\pdfoutput=1

\documentclass[paper=a4, pagesize, english, DIV=calc]{scrartcl}

\usepackage[T1]{fontenc}
\usepackage[ascii]{inputenc}
\usepackage{babel}

\addtokomafont{sectioning}{\rmfamily}
\usepackage{mathptmx,amsfonts}
\usepackage{amsmath, amsthm, mathtools, thmtools, 12many}
\setOTMstyle{dots}

\usepackage{hyperref}

\declaretheorem{theorem}
\declaretheorem[sibling=theorem]{lemma}

\newcommand\NN{\mathbb{N}}
\newcommand\NP{\mathsf{NP}}
\newcommand\PP{\mathsf{P}}
\newcommand\OPT{\mathrm{OPT}}

\title{A PTAS for Scheduling with Tree Assignment Restrictions}

\author{Ulrich M. Schwarz\thanks{%
  ums@informatik.uni-kiel.de}\\
  Institut f\"ur Informatik\\
  Christian-Albrechts-Universit\"at zu Kiel\\
  24098 Kiel, Germany
}

\begin{document}
  \maketitle
  
  \begin{abstract}
    Scheduling with assignment restrictions is an important special case of
    scheduling unrelated machines which has attracted much attention in the
    recent past. While a lower bound on approximability of $3/2$ is known
    for its most general setting, subclasses of the problem admit
    polynomial-time approximation schemes. This note provides a PTAS for
    tree-like hierarchical structures, improving on a recent
    $4/3$-approximation by Huo and Leung~\cite{TheoreticalComputerScience/Huo2010}.
  \end{abstract}
  
  \section{Introduction}
  
  Scheduling on unrelated machines to minimize the makespan is one of the
  classical problem in optimization; here, we are given a set of $n$ jobs
  and $m$ machines, such that execution of a job $j$ on machine $i$ takes
  time $p_{ij}\in\NN$. The objective is to find a schedule, i.e. an
  assignment $\sigma:\ito n\to\ito m$ of the jobs to the machines that
  minimizes the \emph{makespan} 
  $C_{\max}=\max\{\sum_{\sigma(j)=i} p_{ij} :i\in\ito m\}$.
  
  Despite its formal simplicity, it is still not understood completely: no
  approximation result is known that is asymptotically better than the seminal
  $2$-approximation of Lenstra, Shmoys and
  Tardos~\cite{DBLP:journals/mp/LenstraST90}, with asymptotical
  improvements made by Vakhania and
  Shechpin~\cite{DBLP:journals/orl/ShchepinV05}; however, the known lower
  bound on approximability is only $3/2$, also due to Lenstra, Shmoys and
  Tardos.

  A seemingly simpler problem is that of \emph{assignment restrictions}:
  here, for every job $j$ we have a length $p_j\in\NN$ and a set
  $M_j\subset\ito m$ of \emph{feasible machines}, i.e. we have $p_{ij}=p_j$
  for all $i\in M_j$  and $p_{ij}=\infty$ for all $i\not\in M_j$.

  \paragraph{Related results}
  
  As shown already by Lenstra et al.~\cite{DBLP:journals/mp/LenstraST90}, scheduling with arbitrary
  assignment restrictions is also impossible to approximate better than
  within a factor of $3/2$, unless $\PP=\NP$, and for the general case, no
  algorithm better than the $2$-approximation for the unrelated machine case
  is known. However, better results are known for special structures of the
  sets $M_j$. If we have $|M_j|\leq 2$, we can think of jobs as edges in a
  graph whose vertices are the machines, and orienting the edge in one
  direction will increase the load of one of its endpoints. In this \emph{graph
  balancing} setting, Ebenlendr et al.~\cite{DBLP:conf/soda/EbenlendrKS08}
  give a $7/4$-approximation. If the graph additionally a tree, Lee et
  al.~\cite{DBLP:journals/ipl/LeeLP09b} give an FPTAS.
  
  Another type of restriction studied is that of the relation between the
  $M_j$ sets: the most recent results being a PTAS by Muratore
  et al.~\cite{DBLP:journals/orl/MuratoreSW10} for the case of nested
  restrictions, i.e. for each
  two $M_j, M_{j'}$, one of $M_j\subseteq M_{j'}$, $M_j\supseteq M_{j'}$ or
  $M_j\cap M_{j'}=\emptyset$ holds, and a $4/3$-approximation for
  tree-hierarchical
  assignment restrictions by Huo and
  Leung~\cite{TheoreticalComputerScience/Huo2010}. In this setting, again
  machines are considered vertices of a graph, a rooted tree in particular,
  and we impose that the sets $M_j$ must correspond to the machines on a
  path from a node to the root.
  
  For older results, we refer the reader to the
  survey~\cite{IJPE/LeungSurvey2008} by Leung and Li.
  
  \paragraph{Contribution of this note.}
  
  We consider the tree-hierarchical assignment case by Huo and Leung and
  prove the following result:
  \begin{theorem}[label={thm:tree-ptas}]
    Scheduling with tree-hierarchical assignment restrictions admits a PTAS,
    i.e. for every $\epsilon>0$ there is an $(1+O(\epsilon))$-approximation
    with running time polynomial in the input size (but exponential in
    $1/\epsilon$). 
  \end{theorem}

  \section{Rounding and simplifying the instance}

  Our algorithm combines some of the usual techniques for PTAS design such
  as partition into job sizes and geometric rounding with a
  hierarchical dynamic programming approach bottom-up through the tree. In
  this section, we describe the rounding and simplification steps we take to
  make the problem treatable by dynamic programming.

  Throughout the following, let $\epsilon>0$. To simplify the analysis, our
  algorithm will create a solution of length at most $(1+4\epsilon)$ times
  the optimal value $\OPT$. (For simplicity, we use $\OPT$ to refer to both
  an optimal schedule and its makespan, since the distinction is clear from
  context.) 
  Note $\OPT$ must be integral since all jobs lengths are, and it
  is bounded pseudopolynomially in the instance size, for example by
  $\sum_{j=1}^n p_j$. Hence we may, in polynomial time, perform binary
  search over the range of feasible makespans and it is sufficient to give a
  relaxed decision procedure that for a guessed target makespan $C$ 
  yields a schedule of length at most $(1+4\epsilon)C$ whenever a
  schedule of length at most $C$ exists.

  In the following, we call a job \emph{small} if $p_j\leq \epsilon C$,
  otherwise, we call it large. We will round up every large job to be of the
  form $\epsilon C\cdot(1+\epsilon)^k$ for integral $k$. The number $K
  =O(\log_{1+\epsilon} 1/\epsilon)$ of values $k$ that can occur only
  depends on $\epsilon$, i.e. it is a constant for purposes of running time.
  The following classical result holds for this rounding:

  \begin{lemma}
    If there is a schedule of length $C$ of the original instance, 
    there exists a schedule of the rounded instance with a constant number
    $K$ of large job sizes which has length at most $(1+\epsilon)C$.
  \end{lemma}
  
  It is also clear that a feasible schedule of the rounded instance is
  feasible for the original instance by replacing rounded large jobs with
  their (possibly slightly smaller) unrounded counterparts.
  
  We now want to approximately describe every subset of the rounded instance
  by a $(K+1)$-element \emph{configuration tuple}. For large jobs, we simply
  count the number of jobs of each job size, which must be in $\oto n$. For
  small jobs, we  count the total space taken up by them, in integral
  multiples of $\epsilon C$, rounding up. Since every small job has size
  $\leq \epsilon C$, the total size of all small jobs is at most $n\cdot
  \epsilon C$, so this size indicator for small jobs is also from the set
  $\oto n$. In total, the number of configuration tuples is at most
  $(n+1)^{K+1}$, in particular, it is polynomial in the input size.

  We can in this way associate with each node $v$ in the tree the
  configuration tuple $c_v$ of jobs $j$ whose set $M_j$ is the path starting
  in $v$. If $s_v$ is the size multiplicity of the small jobs among them,
  i.e. their total size is in the interval $](s_v-1)\cdot\epsilon C,
  s_v\cdot\epsilon C]$, we add up to one dummy job of size up to $\epsilon
  C$ to make the total size exactly $s_v\cdot\epsilon C$.
  By leaving that job on machine $v$ in the schedule, we obtain
  
  \begin{lemma}
    If there is a schedule of length at most $(1+\epsilon) C$ in the rounded
    instance, there is a schedule of length at most $(1+2\epsilon) C$ in the
    rounded and modified instance.
  \end{lemma}
  
  Let us now consider such a schedule $\sigma$ of length at most $(1+2\epsilon)C$. On
  every machine (node) $v$, a certain subset $\sigma^{-1}(v)$ of jobs is
  scheduled. Hence, it has a corresponding configuration tuple associated
  with it, the total size of which is at most $(1+3\epsilon)C$. The
  additional loss is again incurred because the small jobs in
  $\sigma^{-1}(v)$ might not be an integral multiple of $\epsilon C$.
  It is these configurations that we will find by dynamic programming.

  \section{The algorithm}

  In this section, we describe how to find a feasible assignment of
  configuration tuples to machines, if it exists, and how to convert this
  back into a schedule with a small increase in makespan.
  
  The core of our algorithm is a local procedure which works as follows for
  a node $v$:
  \begin{enumerate}
  \item 
    In the first step, we accumulate the possible subsets of
    not-yet-scheduled jobs that $v$ may need
    to accept from its children. We maintain a set of possible subset
    configuration tuples $S$, which initially contains only the all-zero tuple.
    Then, for each child of $v$ in turn, we consider the set $S'$ of tuples
    it pushes towards the root and set $S:=S+S'=\{c+c': c\in S, c'\in S'\}$. 
    Since the size of $S$ and $S'$ is always polynomial, this can be done in
    polynomial time for every child, and since there are at most $n$
    children, finding the ultimate $S$ with all children taken into account
    also takes polynomial time.

  \item
    Then, we augment $S$ by adding to each tuple the tuple $c_v$ of jobs
    that are only available for scheduling on $v$ and its ancestors. The
    resulting set, which we still denote $S$, still has polynomial size.
  
  \item
    For each $c\in S$, we consider every possible subconfiguration $\hat s$ that 
    can be scheduled on $v$, i.e. is of total size at most $(1+3\epsilon)C$.
    Then, the relative complement $c-\hat c$ corresponds to jobs that would
    need to be pushed towards $v$'s parent node if we schedule according to
    $\hat c$ on $v$. Again, since $S$ is polynomially bounded and the number
    of possible $\hat c$ is as well, this can be done in polynomial time and
    yields a polynomially-sized set of configurations that are possibly
    pushed upwards.
  \end{enumerate}
  
  Our algorithm, for a given target makespan $C$, will execute this
  procedure in any leaf-to-root order, i.e. it is always run on the children
  of a node before it is run on the node itself. We return that a feasible
  schedule exists if it is possible to push up the all-zero configuration
  tuple from the root. The configuration tuples themselves can be obtained
  by standard bookkeeping techniques, i.e. storing, for each
  sum-of-configurations configuration that occurs one (and only one) set of
  witness summands.
  
  Clearly, if there is a feasible assignment of configurations to machines
  of length at most $(1+3\epsilon)C$, the algorithm will find one, too,
  since all configuration tuples that can be pushed into a node are
  considered.
  
  To complete the proof of \autoref{thm:tree-ptas}, it remains to show how
  to assign the jobs. This is trivial for large jobs: we select feasible
  jobs of that size in an arbitrary fashion bottom-up, pushing the remainder
  upwards. Since nothing is pushed beyond the root, all large jobs are
  assigned.
  The situation for small jobs is slightly more complicated, since
  we do not know the exact total size of the small jobs. However, we can
  simply fill the available space in a greedy manner until it is fully used
  (or we run out of small jobs), i.e. the last small job may protude beyond
  the allotted size. Since the last job's size is at most $\epsilon C$ by
  definition, this will increase the makespan of the schedule we generate by
  another $+\epsilon C$ to
  at most $(1+4\epsilon)C$, and it will at most decrease the total size of
  small jobs pushed towards the root, which clearly maintains feasibility of
  the remaining configurations.
  \qed
  
  \section{Conclusion}
  
  This note shows another case, tree-hierarchical structures, in which
  scheduling with assignment restrictions can be approximated within
  arbitrary accuracy. This mostly settles the complexity: an FPTAS cannot
  exist since the setting generalizes the strongly $\NP$-hard problem
  $P||C_{\max}$, the existance of an EPTAS is still open.
  
  For other important settings, the question of inapproximability vs. PTAS 
  is still open: in
  particular, two natural cases would be \emph{cross-free families}, where
  for two sets $M_j, M_{j'}$, $M_j\cup M_{j'}=\ito m$ may also occur in
  addition to the three cases defining nested families as given in the
  introduction, and \emph{interval} restrictions, where every $M_j$ is of
  the form $\{\alpha_j,\dotsc,\omega_j\}$ for a fixed permutation of the
  machines.


\end{document}